\documentclass[letterpaper, 10 pt, conference]{ieeeconf}
\usepackage{graphicx}

\usepackage{algorithm}
\usepackage{algpseudocode}

\usepackage{amsmath}
\usepackage{amssymb}

\usepackage{tabularray}

\usepackage[capitalize]{cleveref}
\crefformat{equation}{(#2#1#3)}
\Crefformat{equation}{Equation~(#2#1#3)}
\Crefname{equation}{Equation}{Eqs.}

\usepackage{amsthm}
\theoremstyle{remark}
\newtheorem{remark}{Remark}
\newtheoremstyle{definition}{}{}{}{}{\bfseries}{.}{.5em}{\thmname{#1}\thmnumber{ #2}\thmnote{ (#3)}}
\theoremstyle{definition}
\newtheorem{definition}{Definition}

\usepackage[font=footnotesize,labelfont=bf]{caption}
\usepackage{caption}
\usepackage{subcaption}

\usepackage{xcolor}

\IEEEoverridecommandlockouts                            
\title{\LARGE \bf
Multi-Robot Task Assignment and Path Finding for \\ Time-Sensitive Missions with Online Task Generation
}

\author{David Thorne and Brett T. Lopez
\thanks{Both authors are with the Verifiable and Control-Theoretic Robotics (VECTR) Laboratory in the Mechanical and Aerospace Engineering Department, University of California Los Angeles, Los Angeles, CA, USA {\tt\small \{davidthorne, btlopez\}@ucla.edu}}%
}

\begin{document}

\maketitle

\thispagestyle{empty}
\pagestyle{empty}

\begin{abstract}

Executing time-sensitive multi-robot missions involves two distinct problems: Multi-Robot Task Assignment (MRTA) and Multi-Agent Path Finding (MAPF). 
Computing safe paths that complete every task and minimize the time to mission completion, or makespan, is a significant computational challenge even for small teams. 
In many missions, tasks can be generated during execution which is typically handled by either recomputing task assignments and paths from scratch, or by modifying existing plans using approximate approaches. 
While performing task reassignment and path finding from scratch produces theoretically optimal results, the computational load makes it too expensive for online implementation.
In this work, we present Time-Sensitive Online Task Assignment and Navigation (TSOTAN), a framework which can quickly incorporate online generated tasks while guaranteeing bounded suboptimal task assignment makespans.
It does this by assessing the quality of partial task reassignments and only performing a complete reoptimization when the makespan exceeds a user specified suboptimality bound. 
Through experiments in 2D environments we demonstrate TSOTAN's ability to produce quality solutions with computation times suitable for online implementation.
\end{abstract}


\section{Introduction}

Time-critical missions such as search-and-rescue, emergency response, warehouse management, and planetary exploration involves teams of robots working together making real-time decisions with information or task priority changing on the fly. 
Task assignment for these types of missions often entails having one or several robots traveling to various locations which, given the need to complete tasks quickly, requires the team to decide the most efficient assignments and corresponding paths. 
This process becomes particularly strenuous when the queue of tasks changes, e.g., specifying a new location of interest to be explored or investigated, which typically requires the complete task assignment and path generation problem be solved for the entire team.
Approximate approaches such as greedy assignment are very fast, but lack any optimality guarantees and can sometimes produce very poor solutions. 
Other algorithms that incorporate new tasks by modifying previous plans without solving the complete problem have been developed \cite{chen2023hybrid}, \cite{forte2021online}, but lack guarantees on the quality of the solution.
This work will present a new method capable of efficiently incorporating online generated tasks with guaranteed performance bounds on time of completion.

   \begin{figure}[t!]
      \centering
      \includegraphics[width=8.5cm]{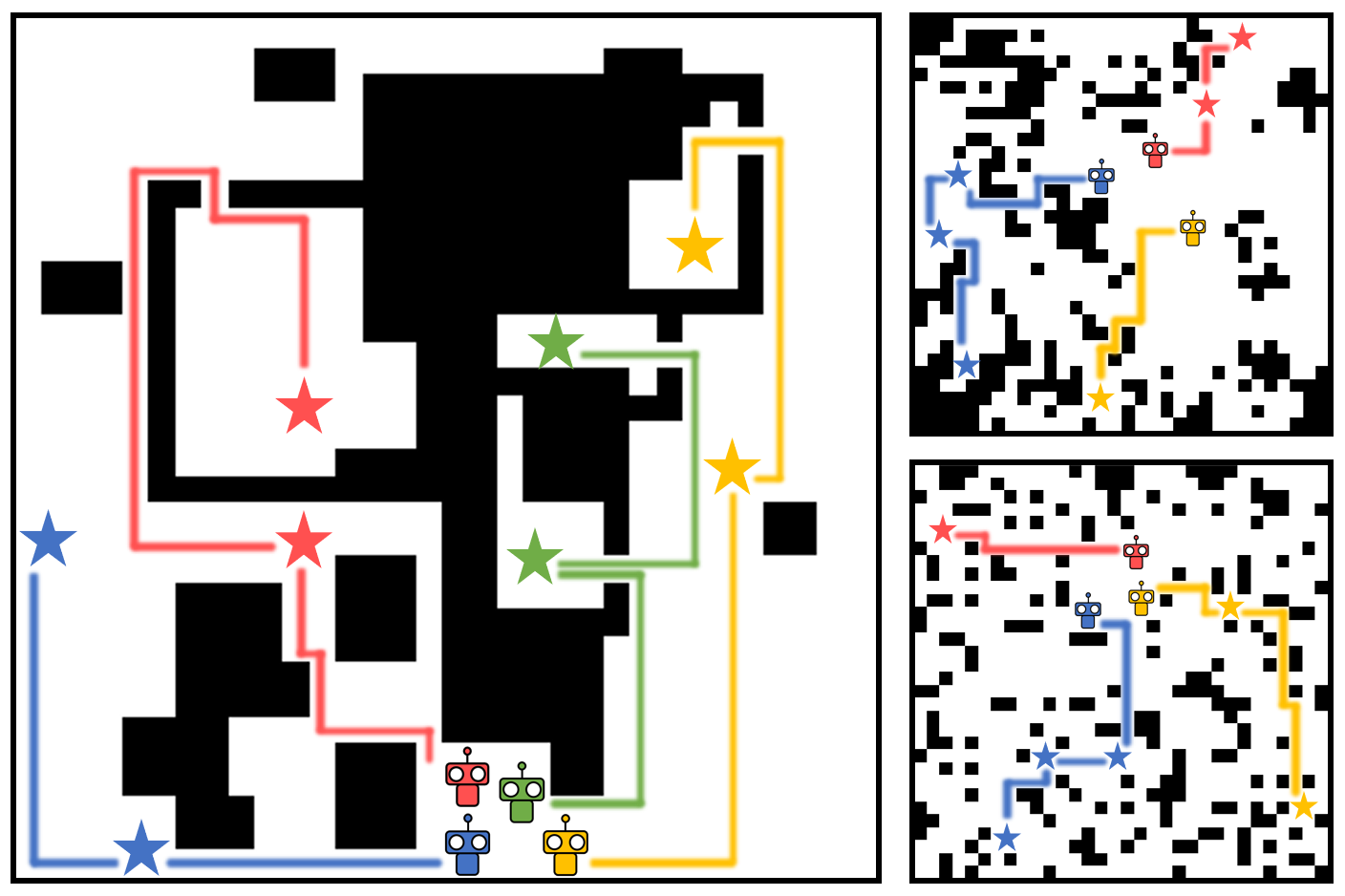}
      \caption{TSOTAN in 2D environments. Multi-robot Task Assignment and Path Finding missions. Left: office, top-right: forest, bottom-right: random (20\% obstacles) environments. Each mission has robots completing tasks on a 32x32 grid made up of open spaces (white) and obstacles (black). The robots (colored in red, green, yellow, and blue) indicate the team start locations, and the stars indicate the task locations.}
      \label{fig:maps}
      \vskip -0.2in
   \end{figure}

Typical approaches to multi-robot coordination can be separated into two sequential problems: Multi-Robot Task Assignment (MRTA), where every task must be assigned as part of a sequence given to a robot, and Multi-Agent Path Finding (MAPF), where collision-free paths to those tasks are generated. 
The MRTA problem requires assigning sets of tasks to robots such that each task is assigned to exactly one robot, and each robot is given a sequence of tasks that specifies a completion order.
Once each robot has an assigned sequence of tasks, any algorithm designed for the classical MAPF problem can be used to generate collision free paths \cite{stern2019multi}. 
The problem with existing methods are 1) the need to compute a complete MRTA solution any time a new task is generated to guarantee optimality and 2) the lack of feedback between MRTA and MAPF algorithms which leads to suboptimal solutions. 
Although algorithms that solve MRTA and MAPF simultaneously exist \cite{honig2018conflict,ren2023cbss,ma2016optimal}, they are not suited for time-sensitive missions where the goal is to minimize the makespan for a potentially large set of tasks.

The main contribution of this paper is Time-Sensitive Online Task Assignment and Navigation (TSOTAN), a coupled MRTA and MAPF framework for multi-robot missions with online task generation. 
TSOTAN guarantees the task assignment makespan will fall within a user specified bound of the optimal makespan without the need to perform computationally expensive task assignment from scratch with each new task.
It uses an initial optimal MRTA solution to track a lower bound on the optimal makespan which is valid for all future timesteps. 
Any time a new task is generated, TSOTAN will attempt a fast but approximate task assignment. 
If the approximate solution exceeds a specified suboptimality bound, it will call for a complete task reassignment which resets the makespan lower bound and ensures computation resources are only invested when needed most. 
Thus, TSOTAN can perform real-time task assignment and path finding for time-sensitive missions while providing strict suboptimality guarantees. 
Additionally, because TSOTAN at its core is an efficient framework for assigning new tasks, it is more robust to failures in hardware and horizon-based path finding algorithms (which can result in deadlocks) as any stuck robot can have their tasks reassigned using the same scheme used for new tasks. 
Simulation results across three environments demonstrate TSOTAN's ability to maintain low computation times while still generating high-quality, near-optimal solutions.

\section{Related Work}

Past research has typically separated MRTA and MAPF into two sequential problems. Using the taxonomy defined by \cite{gerkey2004formal}, the assignment problem is described as a Single-Task robots, Single-Robot tasks, Time-extended Assignment (ST-SR-TA) problem where each task can be accomplished by any robot by travelling to the task location, and all currently known tasks must be taken into account. 
Optimization-based approaches to MRTA include classical algorithms such as the Hungarian algorithm \cite{kuhn1955hungarian}, the Vehicle Routing Problem (VRP) \cite{toth2002vehicle}, and optimization-based formulations derived from the multiple Traveling Salesperson Problem (mTSP) which includes single- and multi-depot variations \cite{kara2006integer,bektas2006multiple,sundar2016generalized,chakraa2023optimization}. 
Mixed-Integer Linear Programming (MILP) formulations are convex when minimizing the sum of all robots' costs, or sum-of-costs, but in order to minimize the makespan, they must employ a bisection method which involves solving the convex problem many times with an additional linear constraint on the makespan \cite{boyd2004convex}. 
Some methods convert mTSP into a single-agent TSP, which means any well-developed TSP solver can be used \cite{oberlin2010today}. 
However, this is not suited to the bisection method, and thus not well suited to minimizing makespan.
Approximate methods for task assignment scale well but lack the performance guarantees of optimization. 
Some approximate approaches include popular heuristic algorithms \cite{helsgaun2009general}, region assignment \cite{schneider1998territorial}, and learning-based assignment \cite{park2021cooperative}. 
Most relevant to this work is iterative approaches that look to improve suboptimal solutions \cite{chen2023hybrid}, \cite{zheng2009k}. 
While iterative methods provide explicit assignments and are better suited to incorporating new tasks, they cannot guarantee optimal solutions and are designed to reduce the sum-of-costs.

The classical (or non-anonymous) MAPF problem has received significant interest recently for its applications in warehouses where algorithms need to efficiently compute safe paths for many agents with predefined destinations \cite{stern2019multi}. 
Conflict-Based Search \cite{sharon2015conflict} is a well-studied optimal algorithm with many variations that improve its computation time \cite{barer2014suboptimal,li2021lifelong}. 
Receding-Horizon Conflict Resolution (RHCR) \cite{li2021lifelong} is one such variation that improves computation time by solving MAPF in a windowed manner, but its horizon-based planning can result in deadlocks. 
Some MAPF algorithms incorporate windowed replanning and aim to improve performance in dense environments \cite{arul2022dense,wu2020multi}, but these approaches are designed to improve suboptimal path finding results that arise from decentralized planning.

While most approaches solve MRTA and MAPF separately, several algorithms have been proposed that solve them simultaneously to improve the overall performance.
MRTA and MAPF can be solved simultaneously with a forest search algorithm \cite{honig2018conflict}, \cite{ren2023cbss}, \cite{wagner2012subdimensional}, but this method is not well suited to minimizing makespan as it requires solving MRTA quickly and repeatedly. 
The anonymous MAPF problem is equivalent to simultaneous MRTA and MAPF with instantaneous assignment where each robot is given one task at a time. 
Several algorithms have been proposed for anonymous MAPF minimizing makespan \cite{forte2021online}, \cite{turpin2014goal}, with multiple algorithms running in polynomial time \cite{ma2016optimal}, \cite{okumura2023solving}; however, they are limited by only being able to assign one task to each robot at a time. 

\section{Problem Formulation}

    \begin{figure*}[thpb]
          \vskip 0.06in
          \centering
          \includegraphics[width=.98\textwidth,height=4cm]{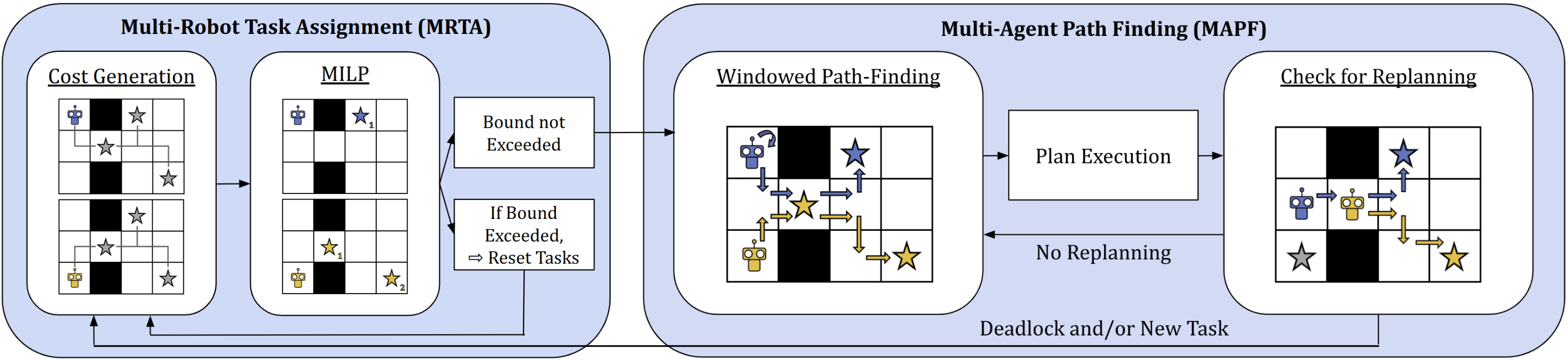}
          \caption{TSOTAN block diagram. Task assignment is comprised of cost generation, MILP optimization, and a solution bound check. Path finding includes the windowed path finding algorithm and replan check at the end of each execution horizon.}
          \label{fig:block_diagram}
          \vskip -0.2in
    \end{figure*}

The combined MRTA and MAPF problem is to assign tasks to robots and generate safe paths to those tasks while minimizing the makespan, which is the time to complete all tasks (to be defined mathematically below). Let the set of $N$ agents be denoted by index $i \in I=\{1,2,\dots,N\}$, and the set of $M$ tasks by index $\tau \in T =\{1,2,\dots,M\}$ where $M$ can increase as tasks are generated during the mission. 
A superscript ${i}$ will be used to denote an element belonging to agent $i$ unless otherwise noted. 
Each agent stores their own undirected graph $G^{i}=(V^{i},E^{i})$ where the vertex set $V^{i}$ represents every allowable location for an agent and the edge set $E^{i} \in V^{i}\times V^{i}$ denotes the set of all possible actions the agent can take to move between vertices in $V^{i}$. 
An edge connecting vertices $v_{a}$ and $v_{b}$ is denoted as $e_{a,b}$.
Each task can be assigned to only one agent, but agents can be assigned an arbitrary number of tasks with an intended sequence of completion. 
A MRTA solution is called a task assignment and is denoted as $\Pi$, where $\Pi(i) = \{\tau^{i}_{1},\tau^{i}_{2},...\}$ is the set of tasks assigned to agent $i$. 
A task is completed when a robot arrives at the first incomplete task in its assignment.
We will make use of the following definitions in the sequel.

\begin{definition}
    A \emph{path}, denoted as $\pi(v_{0},v_{f})= \pi_{0,f} = \{[v_{0},t_{0}],[v_{1},t_{1}],...,[v_{f},t_{f}]\}$, represents the ordered set of vertices and associated timesteps which a robot follows to travel from $v_{0}$ (initial state) to $v_{f}$ (final state). 
\end{definition}

\begin{definition}
    A \emph{safe} path is one that guarantees no collisions with the environment or other robots.
\end{definition}

\begin{definition}
    A \emph{solution} to the combined task assignment and path finding problem is a set of safe paths for all agents in which every available task is completed.
\end{definition}

\section{Preliminaries: Plan Modifications}
For each path we associate a cost $\rho(\pi)$ which can represent any non-negative metric such as distance, energy spent, or time. 
We define the path cost and makespan to depend on time only, e.g., 
\begin{align}
\label{eq:path_cost}
\rho(\pi_{0,f}^{i}) &= \sum_{n=0}^{f-1} e_{n,n+1}/r^{i}_{s}\\
\label{eq:makespan_def}
J &= \min_{i}(\rho(\pi^{i}))
\end{align}
where the value of $e_{n,n+1}$ is the distance of the edge between $v_{n}$ and $v_{n+1}$, and $r^{i}_{s}$ is the speed of robot $i$ which is assumed to be either constant or parameterized by environmental factors, e.g., terrain. 
A path that minimizes the cost between any two vertices is called a \emph{direct} path. The cost of a direct path $\rho(\pi_{a,b})$ is always non-negative and satisfies $\rho(\pi_{a,c}) \leq \rho(\pi_{a,b})+\rho(\pi_{b,c})$. 
We enforce that every robot's path must be composed of one direct path segment for each task in that robot's task assignment.
Two paths can be appended if they share a similar start/end point, i.e., $\pi_{a,b}+\pi_{b,c}=\pi_{a,c}$, and the combined cost is the sum of the component paths' costs $\rho(\pi_{a,c})=\rho(\pi_{a,b})+\rho(\pi_{b,c})$. 
A path can also be split into multiple components where the sum of the cost of each of the components equals the cost of the original path. 

Through a combination of appending and splitting path segments, a new task can be inserted into an existing path. 
This is done by splitting the original path into three segments, removing the middle section, and adding two new direct paths to and from the task. 
These steps, the resulting path cost, and the change in path cost are given in \cref{eq:path_mod}-\cref{eq:path_cost_change} where $\pi^*_{0,f}$ denotes the new path:
\begin{align}
\label{eq:path_mod}
\pi_{0,f}&=\pi_{0,a}+\pi_{a,c}+\pi_{c,f}\\
\label{eq:path_mod2}
\pi^{*}_{0,f}&=\pi_{0,a}+\pi_{a,b}+\pi_{b,c}+\pi_{c,f}\\
\label{eq:path_mod3}
\rho(\pi^{*}_{0,f})&=\rho(\pi_{0,a})+\rho(\pi_{a,b})+\rho(\pi_{b,c})+\rho(\pi_{c,f})\\
\label{eq:path_cost_change}
\rho(\pi^{*}_{0,f})-&\rho(\pi_{0,f}) = \rho(\pi_{a,b})+\rho(\pi_{b,c})-\rho(\pi_{a,c}) \geq 0.
\end{align}

The change in cost is the additional cost of the paths to and from the new position, minus the deleted path cost. 
If the paths removed and added are direct paths, this action of inserting a new task follows the triangle inequality and the change in cost must be non-negative as shown in \cref{eq:path_cost_change}.
Given that adding a new task will never decrease any robot's path cost, it follows that adding a new task will never decrease the makespan of any previous assignment.

\begin{remark}
    \label{remark:remark2}
    Knowing how path costs change by adding tasks means we can set and track bounds on the optimal makespan throughout the mission. We will use a tracked lower bound on the optimal makespan later to assess the quality of approximate task assignment solutions.
    For example, if a solution is optimal at time $T$ with makespan $J$, $\Omega(t) = J-(t-T)$ $\forall t \geq T$ is a valid lower bound on the optimal makespan at any time $t$ regardless of the used assignment and robot paths or additional tasks generated. This is true as $(t-T)$ is the amount of time since the optimal solution was found, and the optimal makespan cannot decrease by more than the amount of time passed. 
\end{remark}

\section{Methods}
TSOTAN is a loosely coupled MRTA and MAPF framework which uses a mTSP optimization to form the initial task assignment and makespan lower bound. 
Collision-free paths are generated by a horizon-based path finding algorithm, where at the end of each execution window the algorithm will check for deadlocked robots or newly generated tasks which merit partial or complete task reassignment. 
The framework is shown as a block diagram connecting four critical algorithms in \cref{fig:block_diagram}.

\subsection{Cost Generation}

\label{cost_gen}
The first step in TSOTAN's pipeline is assigning all avialable tasks to the team. 
The set of tasks to be assigned is denoted by $\mathcal{U}$, which is a subset of the set of all incomplete tasks $\mathcal{T}$.
Before any MRTA optimization, each robot must build a square cost matrix $C \in \mathbb{R}^{(M+1) \times (M+1)}$ by computing the cost to travel between every task in $\mathcal{U}$ as well as a specified starting location.
An entry $c^{i}_{j,k}$ represents the cost for robot $i$ to travel from position $j$ to $k$, where the index 0 represents the starting location of the robot.
A \emph{complete} task assignment is when $\mathcal{U}=\mathcal{T}$, but the more general case when some task have already been assigned ($\mathcal{U} \subseteq \mathcal{T}$) is called partial task assignment. 
Partial task assignment is a faster, approximate version of complete task assignment in which all tasks in $\mathcal{U}$ must be assigned without changing the previous assignment.
During cost generation any robot with an existing assignment makes a \emph{modified} cost matrix where the starting position is the position of the last assigned task, and the remaining path cost is added to the first row of $C^{i}$.
The classical mTSP requires all robots' paths to start and end at a home depot, but we want robots to end the mission at their final task location. To maintain the mTSP formulation but ignore the return to home requirement, the first column of each cost matrix $c^{i}_{j0}$ is set to zero and the final action in each robot's assigned sequence is ignored.

\subsection{Task Assignment via MILP Optimization}

We designed an algorithm (shown in Algorithm 1) for both partial and complete task assignment that only requires a set of cost matrices as described in \cref{cost_gen}, which can be sped up with the input of previously known makespan bounds. 
When robots have already been assigned tasks, the algorithm also requires the previous task assignment as input. 
It uses an MILP formulation of the multiple Traveling Salesperson Problem (mTSP) with multiple depots and no set final destinations. 
Since the goal is to minimize the makespan, the formulation must perform a min-max optimization on the robot path costs. 
To achieve this, we use the convex formulation for multi-depot mTSP given in \cite{kara2006integer} that minimizes the sum of all robot path costs with two modifications. 
For the first modification, we turn the optimization into a min-max operation by introducing a new variable constraint $p$ that limits each robot's path cost $\sum C^{i} \circ X^{i} \leq p \quad \forall i \in I$. 
The variable $p$ will converge on the optimal makespan through a binary search until the bounds become smaller than a threshold $\Gamma$ based on the success or failure of a sum-of-costs optimization. 
This is similar to the bisection method described in Algorithm 4.1 of Boyd and Vandenberghe \cite{boyd2004convex}. 
The second modification is to omit some constraints given in \cite{kara2006integer} that set min and max limits on the number of tasks visited by each robot. 
The convex sum-of-costs optimization is called $mTSP$ in Algorithm 1, and it is implemented using GUROBI \cite{gurobi}.

We use $\omega$ to denote a makespan lower bound, $O$ to denote an upper bound, and $J$ to denote the makespan. 
If no lower bound is given, $\omega$ is set as the $M^{\mathrm{th}}$ cheapest nonzero cost amongst all cost matrices. If no upper bound is given, we set $O$ as the makespan of the sum-of-costs optimal task assignment.
The threshold $\Gamma$ can be set to the smallest possible change in makespan to ensure an optimal solution, but in practice we used $\Gamma = \omega*(\mu-1) \; \forall \: \mu > 0$, which will return a makespan bounded by $\mu * J_{opt}$. This is easily proved as the largest makespan the binary search in Algorithm 1 will return is $J_{max}=O=\mu*\omega$. 
Since by definition $\omega \leq J_{opt}$, it follows that $J_{max} = \mu*\omega \leq \mu *J_{opt}$.

During partial reassignment, the binary search will not swap tasks between robots or change assignment orders as that would require the complete reassignment of all tasks. 
Instead, the initial binary search in Algorithm 1 will add all tasks in $\mathcal{U}$ to the end of any previous task assignments. 
In an attempt to improve partial reassignment, the algorithm will find any robots with previous assignments that received new tasks and reoptimize those robot's task sequences as a single-agent TSP optimization.
Although partial reassignment does not guarantee optimality, it is significantly faster than a complete reoptimization. Using the tracked lower bound on makespan $\Omega$, it can be determined if the suboptimal solution produced by partial reassignment (denoted by $O$ here) is within an acceptable tolerance defined by the parameter $\gamma$. 
If $O > \gamma \Omega$, then the algorithm will force a complete optimization with all incomplete tasks and robots in their current positions. 
$O$ is a valid upper bound on the optimal makespan for the complete MRTA optimization, so Algorithm 1 can use the initial tight bounds $\Omega \leq J_{opt} \leq O$.

\begin{figure}[!t]
    \vskip 0.06in
    \centering
     \includegraphics[width=.99\columnwidth]{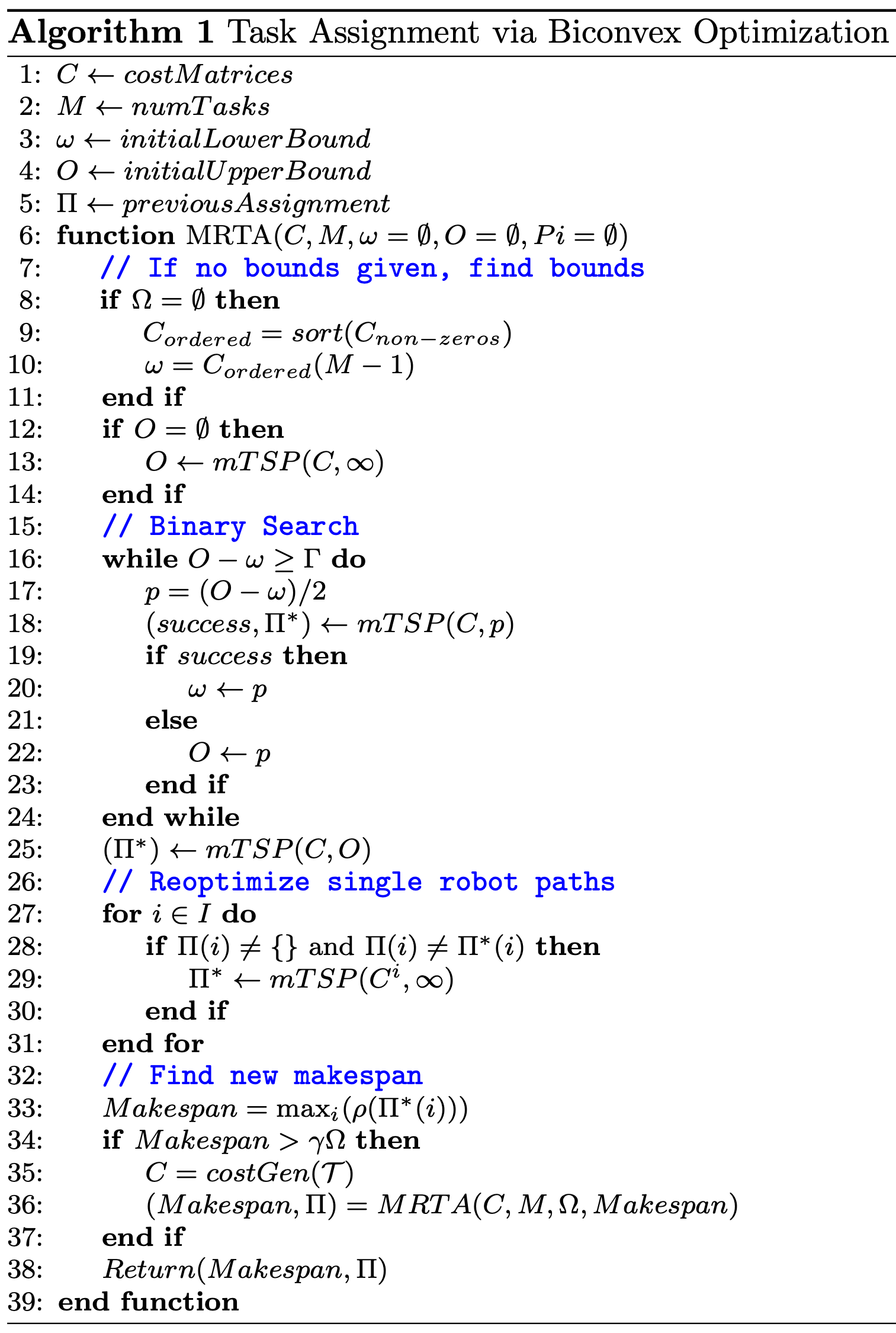}
     \vskip -0.2in
\end{figure}

\begin{figure*}[h]
  \centering
  \vskip 0.06in
  \includegraphics[width=15cm]{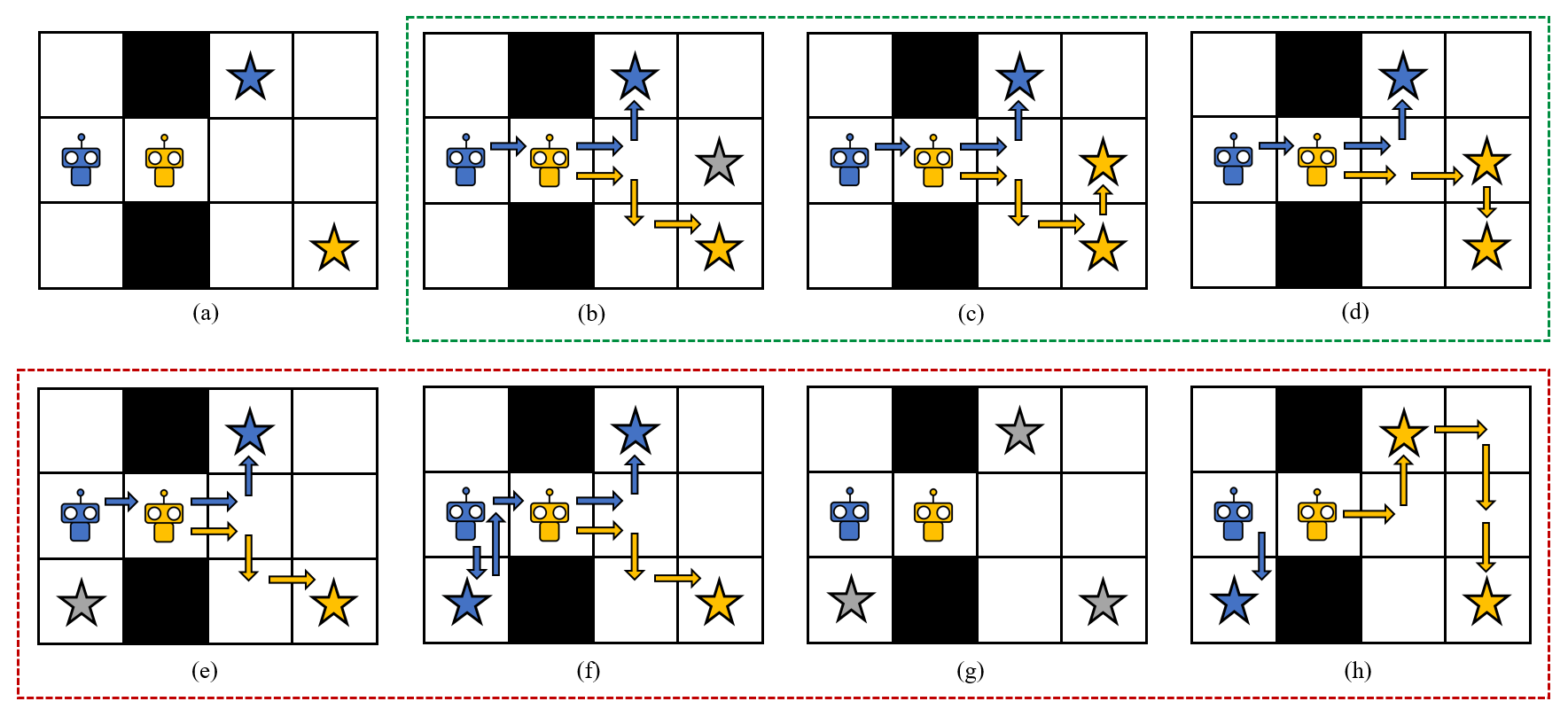}
  \vskip -0.1in
  \caption{Task reassignment examples. (a) Initial problem state, $\Omega=3$, $\gamma = 1.5$. (b) First scenario (denoted by the green bounding box): a new task (grey star) is generated at [1,3]. (c) After the initial reassignment, the yellow agent attempts to re-optimize its path. (d) The new makespan $J=3$ is the same as the previous makespan, so reassignment is complete. (e) Second scenario (denoted by the red bounding box): a new task is generated at [2,0]. (f) After the initial reassignment and re-optimization, the new makespan $J=5 > \gamma \Omega = 4.5$ exceeds suboptimal bound so full task reassignment is required. (g) Full task assignment with initial bounds on the makespan of $3 \leq J \leq 5$. (h) After full task reassignment the new makespan is $J=5$. Note that the blue agent could have been given two tasks so $J=5$ but the solution with lower sum-of-costs is always chosen.}
  \label{fig:case_study}
  \vskip -0.2in
\end{figure*}

\subsection{Windowed Path Finding}

After task assignment, a set of collision-free paths must be computed that safely navigate each robot to its assigned tasks as fast as possible. 
Implementing a windowed MAPF algorithm makes the overall computation cheaper while naturally leading to planning horizons where TSOTAN can evaluate the state of the mission. 
For this purpose we implement Receding-Horizon Collision Resolution (RHCR) \cite{li2021lifelong}, which was developed as a suboptimal, horizon-based variant of the popular Conflict-Based Search (CBS) \cite{sharon2015conflict} algorithm. 
CBS and RHCR both work as two-level search algorithms where the low-level can plan constrained paths on a grid, and the high-level is a tree search that finds and resolves potential conflicts by generating new nodes with additional constraints. We make two notable modifications to RHCR in how conflicts are defined and how to address deadlock.

MAPF algorithms typically work on shared graphs and define conflicts as two robots sharing the same node or edge at a given timestep \cite{stern2019multi}. 
We define conflicts in continuous space to facilitate each robot having their own unique undirected graph $G^{i}=(V^{i},E^{i})$. Instead of node conflicts, we assign each robot a collision radius and ensure no two robots' collision spheres intersect at any timestep. 
Instead of edge conflicts, we generate a collision tunnel between timesteps and ensure that no two collision tunnels intersect, which also enables robots that can move across multiple edges in one timestep as collision tunnels can be stitched together.

Windowed MAPF can occasionally lead to deadlock where two agents block each other but neither can plan far enough ahead to take an more expensive open path. 
The authors of \cite{li2021lifelong} suggest increasing the planning horizon when deadlocks are detected, but this increases the computational cost and defeats the purpose of implementing a windowed path finding algorithm in the first place. 
Instead, we reassign the tasks of robots in deadlock and broadcast those robots' positions to the rest of the team as occupied space for the purposes of cost generation. 
Because this leads to a theoretically worse makespan during task assignment, any tracked lower bound on the optimal makespan remains valid.

\subsection{Replanning Check}

At the end of each path finding execution window, TSOTAN will update the lower bound $\Omega$ and assess the state of the mission. 
Assuming perfect execution, the optimal makespan would decrease by $T_{e}$, so we update $\Omega$ such that $\Omega \leftarrow \Omega - T_{e}$. 
Whenever a new task is generated it is automatically added to $\mathcal{U}$ and $\mathcal{T}$. 
Detecting deadlocks is a challenging problem with its own field of research, we use a simple detection scheme for our simulations that detects deadlocks if the entire team does not make sufficient progress towards their goals during an execution window. 
If any robots are identified as deadlocked, we unassign their tasks, add those tasks back into $\mathcal{U}$, and broadcast their positions for the rest of the team to use in cost generation. 
If, after deadlock detection, $size(\mathcal{U}) \neq 0$, the replanning check will call for a partial task reassignment before continuing with the mission.

\subsection{TSOTAN: Illustrative Example}
To demonstrate TSOTAN, two scenarios are presented. For this example, we assume an execution horizon $T_{e} = 2$, a solution tolerance $\gamma = 1.5$ and tracked makespan lower bound $\Omega=3$.
\cref{fig:case_study}a shows the initial state of the problem. In the first row (3b-3d), a new task is generated and partial reassignment gives this task to the yellow robot. 
However, as this robot already had assigned tasks and was given a new task, a single-agent TSP optimization resequences the yellow agent's assignment and the new solution makespan is 3. 
As the new makespan is less than $\gamma \Omega$, task assignment is done.

In the second row (3e-3h), a new task is generated and partial reassignment gives this task to the blue robot. 
Even after calling the single-agent TSP to optimize its path, the makespan increases to 5. 
Since $\gamma \Omega=4.5$, a complete task assignment will be triggered. 
Using the improved bounds $\Omega=3$ and $O=5$, the optimization finds the new optimal solution with a makespan $J=5$ and we reset $\Omega=5$. 
Two possible assignments could have led to $J=5$, but the one with the lower sum-of-costs was chosen because the optimization will minimize the sum-of-costs under the makespan constraint.

\section{Results}

\begin{figure}[t!]
     \vskip 0.06in
     \hskip 0.05in
     \centering
     \begin{subfigure}[b]{4.1cm}
         \centering
         \includegraphics[width=4.1cm]{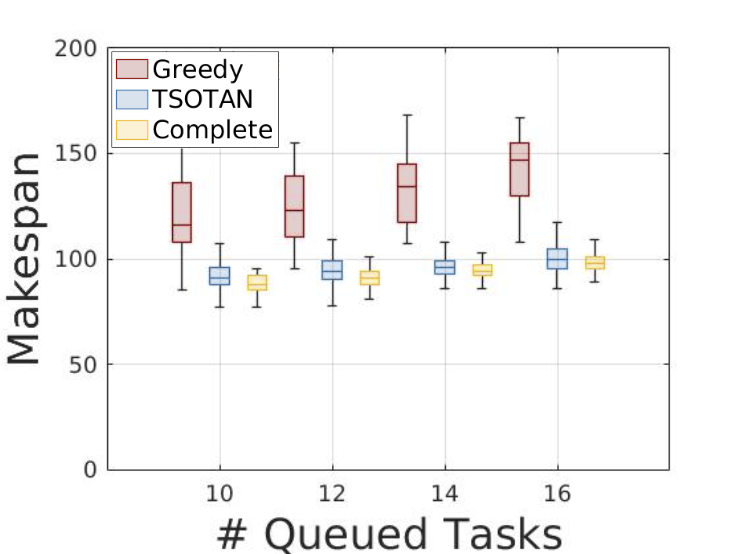}
         \caption{Makespan for office.}
         \label{fig:office-makespan}
     \end{subfigure}
     \hfill
     \begin{subfigure}[b]{4.1cm}
         \centering
         \includegraphics[width=4.1cm]{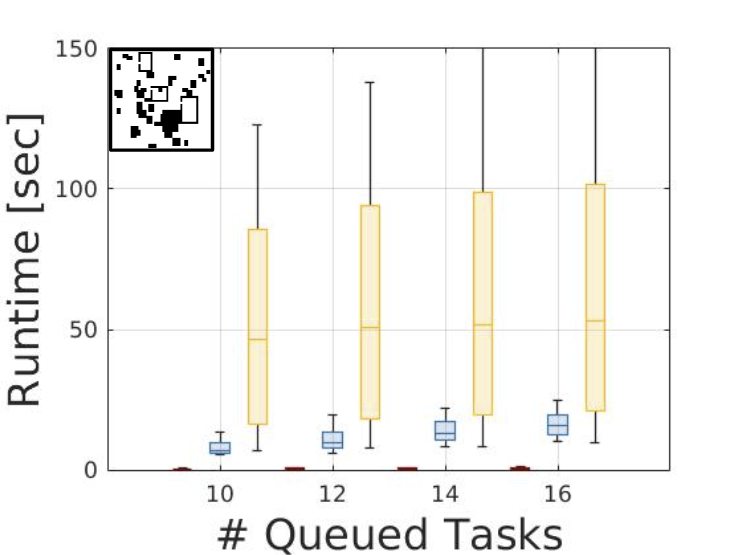}
         \caption{Runtime for office.}
         \label{fig:office-runtime}
     \end{subfigure}\\
     \hskip 0.05in
     \begin{subfigure}[b]{4.1cm}
         \centering
         \includegraphics[width=4.1cm]{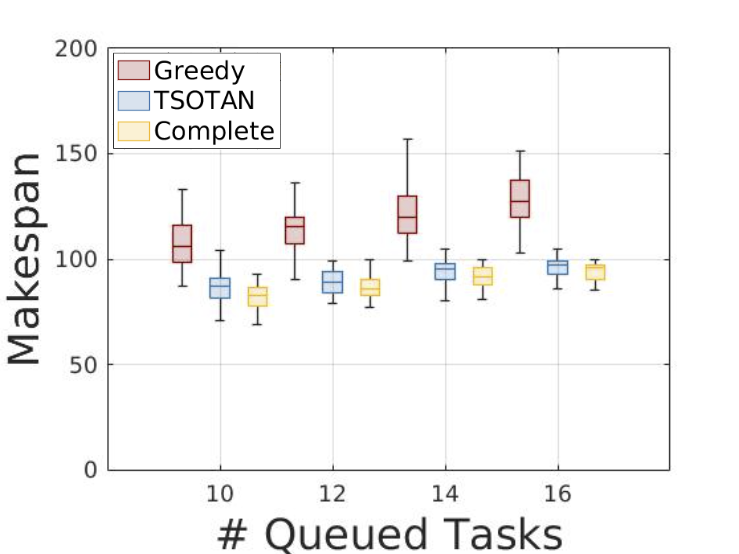}
         \caption{Makespan for forest.}
         \label{fig:forest-makespace}
     \end{subfigure}
     \hfill
     \begin{subfigure}[b]{4.1cm}
         \centering
         \includegraphics[width=4.1cm]{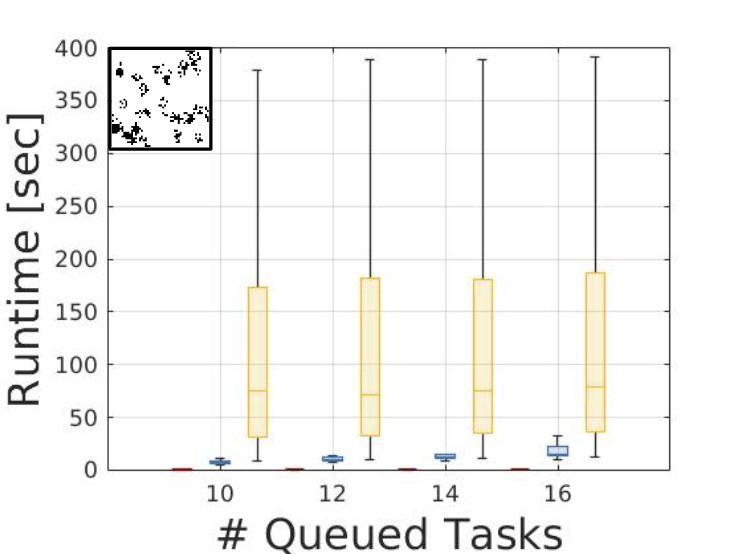}
         \caption{Runtime for forest.}
         \label{fig:forest-runtime}
     \end{subfigure}\\
     \hskip 0.05in
     \begin{subfigure}[b]{4.1cm}
         \centering
         \includegraphics[width=4.1cm]{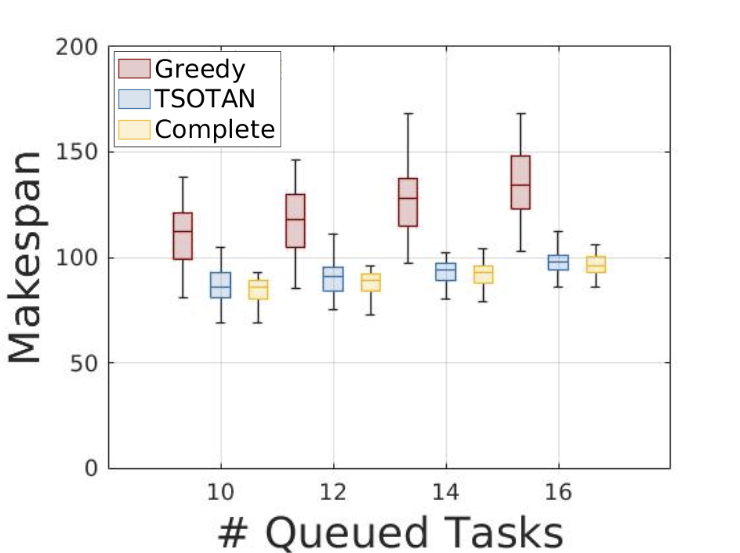}
         \caption{Makespan for random.}
         \label{fig:random-makespan}
     \end{subfigure}
     \hfill
     \begin{subfigure}[b]{4.1cm}
         \centering
         \includegraphics[width=4.1cm]{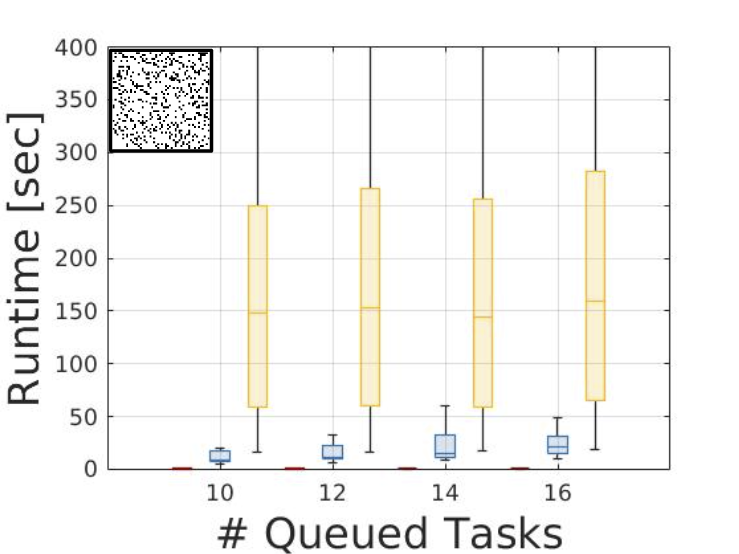}
         \caption{Runtime for random.}
         \label{fig:random-runtime}
     \end{subfigure}\\
        \label{fig:naive_comp}
        \caption{Makespan and runtime comaprison. Left column shows makespan in timesteps across three environment types: office (a), forest (c), and random obstacles (20\%) (e). Right column shows computation time after the initial MRTA solution in seconds across the same environment types. All environments are $50 \times 50$ grid worlds with 6 robots and 10 initial tasks.}
    \label{fig:results-1}
     \vskip -0.2in
\end{figure}

We compared TSOTAN against two naive methods which represent two extremes in the trade-off between computation time and optimality in online task assignment. 
The greedy method assigns tasks one at a time to the end of a robot's assignment such that the makespan is minimized. 
The complete method performs complete MRTA optimization using $\mathcal{T}$ any time task reassignment is called. 
All three methods use the same optimal solution to the initial MRTA problem using 6 robots and 10 initial tasks, which is why runtime statistics do not include the initial MRTA solve time. 
Simulations were run in three distinct 2D environment types (office, forest, random), examples of which can be seen in \cref{fig:maps}. 
Queued tasks is the number of tasks that will be generated online, where the probability of a task being generated at any timestep is 25\%. 
The makespan tolerance is set at $\gamma=1.5$. 
For each method, 50 simulations with a cutoff time of 600s were conducted given the map and number of queued tasks. 

\paragraph{Performance Evaluation} 
TSOTAN outperforms both naive methods by producing quality solutions in shorter runtimes. 
TSOTAN had an average makespan of 93.09, which is only a 2.8\% increase compared to 90.56 for the complete method, but a significant 24.5\% decrease compared to 123.33 for the greedy method.
The largest increase in average makespan compared to the complete method was 5.6\% recorded in the office map with 10 queued tasks.
Makespans can be seen in \cref{fig:office-makespan,fig:forest-makespace,fig:random-makespan}.
TSOTAN's average runtime was 33.39 seconds, which is several times larger than the greedy method's average of 0.74 seconds but, as stated above, produces solutions with much lower makespan.
The complete method had an average runtime of 115.41 seconds which is impractical for real-time applications. 
Runtimes can be seen in \cref{fig:office-runtime,fig:forest-runtime,fig:random-runtime}. 
TSOTAN timed out in 2\% of office missions, 3\% of forest missions, and 2.5\% of random missions. 
The complete method timed out in 12\% of office missions, 22\% of all forest missions, and 22.5\% of random missions. The greedy method never timed out.

\begin{figure}[t!]
     \vskip 0.06in
     \centering
     \label{fig:ablate}
     \begin{subfigure}[b]{4.1cm}
         \centering
         \includegraphics[width=4.1cm]{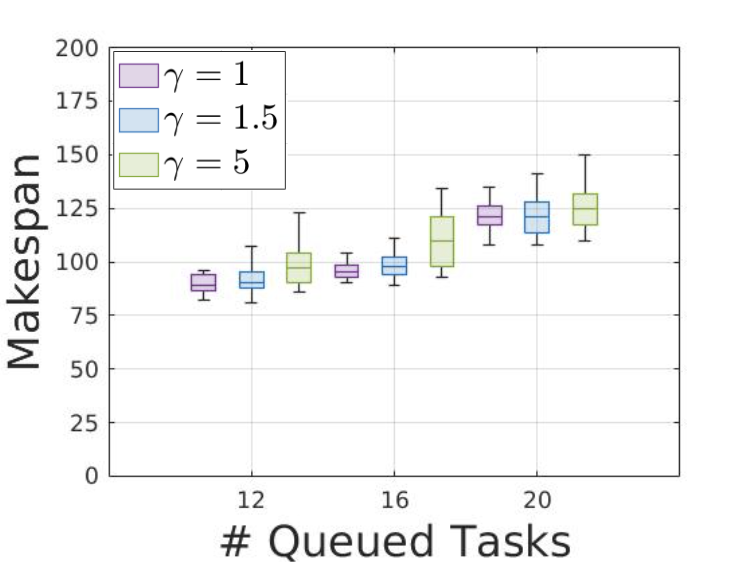}
         \caption{Makespan vs. queued tasks.}
         \label{fig:abl_a}
     \end{subfigure}
     \begin{subfigure}[b]{4.1cm}
         \centering
         \includegraphics[width=4.1cm]{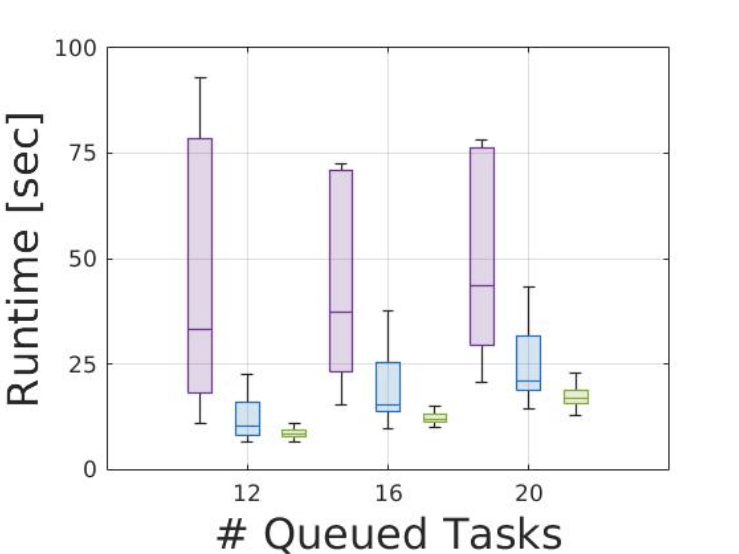}
         \caption{Runtime vs. queued tasks.}
         \label{fig:abl_b}
     \end{subfigure}
     \caption{MRTA Tolerance Ablation Study. (left) Makespan vs. number of queued tasks and (right) computation time after initial task assignment vs. number of queued tasks for three MRTA tolerance values.}
     \label{fig:results-2}
     \vskip -0.2in
\end{figure}

\paragraph{Ablation Study} 
Balancing computation time versus performance is problem specific, but the value of $\gamma$ is meant to give operators a parameter to decide how to strike that balance. 
\Cref{fig:results-2} shows the results of TSOTAN with three values of $\gamma$, where $\gamma=1$ produces the same solutions as the complete method in return for longer computation times, and $\gamma=5$ behaves more like the greedy method with shorter computation times but worse makespans. 
The results in Fig. 5 were collected over 25 simulations for each number of queued tasks on the $50 \times 50$ random map with 20\% obstacles.
The average makespan across all simulations was 98.7 for $\gamma=1$, 100.48 for $\gamma=1.5$, and 107.95 for $\gamma=5$. The largest difference was a 15.8\% increase in average makespan between $\gamma=1$ and $\gamma=5$ with 16 queued tasks, but the difference was only a 10.5\% increase in average makespan between $\gamma=1$ and $\gamma=5$ with 20 queued tasks.
The average runtime across all simulations was 91.32 seconds for $\gamma=1$, 44.55 seconds for $\gamma=1.5$, and 12.48 seconds for $\gamma=5$, where 32\% of the solutions using $\gamma=1$ timed out after 300 seconds.
Although this does not give a full characterization of the trade-off between computation time and solution quality as a function of $\gamma$, these results suggest that increasing the suboptimality bound from $\gamma=1$ to $\gamma=5$ cuts the computation time by an order of magnitude while only losing 10-20\% solution quality.

\section{Conclusion}

This work presented TSOTAN, a framework which can quickly incorporate online generated tasks while guaranteeing bounded suboptimal task assignment makespans. 
We demonstrated through 2D simulations that TSOTAN produces high quality solutions using an order of magnitude less computation time compared to a complete reassignment approach. 
Further, we investigated the relationship between solution makespan and computation time while varying the MRTA solution tolerance. 
The results suggest that by specifying a single parameter, TSOTAN gives the flexibility to balance computation time versus solution quality for any environment or mission type. 
Future work will further characterize the trade-off between performance and computation time with MRTA tolerance in various environments, as well as look to implement TSOTAN in laboratory experiments.

\vspace{0.1in}
\noindent \textbf{Acknowledgements:} The authors thank Jacob Sayono for his help with making figures and simulation environments.

\bibliographystyle{IEEEtran}
\bibliography{root}

\end{document}